\PassOptionsToPackage{bookmarks=false,colorlinks=true,citecolor=red}{hyperref}

\documentclass[12pt,pdflatex,sn-nature,iicol]{sn-jnl}

\usepackage{cuted}

\newenvironment{widetext}
    {\begin{strip}}
    {\end{strip}}

\usepackage{amsmath}
\usepackage{mathtools}
\usepackage{amssymb}
\usepackage[utf8]{inputenc}
\usepackage{physics}
\usepackage{tabularx}
\usepackage{overpic}
\usepackage{balance}
\usepackage{color}
\usepackage{xcolor}
\usepackage{graphicx}
\usepackage{import}
\usepackage{tabularx}
\usepackage{dcolumn} 
\usepackage{bm}      
\usepackage{array}   
\usepackage{amsfonts}
\usepackage{tikz}
\usepackage[T1]{fontenc}
\usepackage{tabularray}[2021-06-01]

\usepackage[T1]{fontenc}

\usepackage[english]{babel}

\usepackage[utf8]{inputenc}
\usepackage{array}
\usepackage[T1]{fontenc}
\usepackage{mathptmx}

\include{MyCommand}

\DeclareUnicodeCharacter{2212}{-}

\makeatletter
\newcommand*{\addFileDependency}[1]{
  \typeout{(#1)}
  \@addtofilelist{#1}
  \IfFileExists{#1}{}{\typeout{No file #1.}}
}
\makeatother

\include{MyCommand}

\begin{document} 


\title{Resonance-induced quantum non-Gaussian states below criticality} %




\author{Daniel~J.~ Nader}
\email{daniel.juliannader@upol.cz}
 \affil{Department of Optics, Faculty of Science, Palack\'y University, Olomouc 77900, Czech Republic}

\author{Pavel Str\'ansk\'y}
 \affil{
 Institute of Particle and Nuclear Physics, Faculty of Mathematics and Physics,
Charles University, V Hole\v{s}ovi\v{c}k\'ach 2, 18000 Prague, Czech Republic}
 
 \author{Jorge Ch\'avez-Carlos}
 \affil{Physics Department, Cinvestav, AP 14-740, Mexico City 07000, Mexico}



\author{Pavel Cejnar}
\email{pavel.cejnar@matfyz.cuni.cz}
 \affil{
 Institute of Particle and Nuclear Physics, Faculty of Mathematics and Physics,
Charles University, V Hole\v{s}ovi\v{c}k\'ach 2, 18000 Prague, Czech Republic}
  \author{Radim Filip}
   \email{filip@optics.upol.cz} 
\affil{Department of Optics, Faculty of Science, Palack\'y University, Olomouc 77900, Czech Republic}
 

\begin{abstract}
TThe Quantum Rabi Model (QRM), consisting of a single qubit coupled to a harmonic oscillator, provides a paradigmatic platform for exploring critical phenomena investigated for their applications in quantum technologies. In the superradiant phase of the QRM with coupling strength above the critical point, the bosonic ground state exhibits temporally stable quantum non-Gaussian features. However, the quantum non-Gaussian features of such superradiant ground states can be fragile against loss during preparation. Here, we show that persistent 
quantum non-Gaussian features can instead be realized below the critical point through weak periodic qubit modulation. We analyze an adiabatic ramp protocol in which the modulation amplitude is gradually increased, allowing the system to approximately follow the instantaneous Floquet state. When the modulation period is tuned close to Poincaré–Birkhoff resonances, the resulting state exhibits pronounced quantum non-Gaussian features in the directly measurable Wigner function 
with a complex phase-space structure that is considerably more robust against loss than that of the superradiant ground state.
The protocol of adiabatic ramps provides an attractive route for generating long-lived
quantum non-Gaussian effects below the critical point.
\end{abstract}


\maketitle

\begin{figure*}[!htbp]
    \centering
    \includegraphics[width=\textwidth]{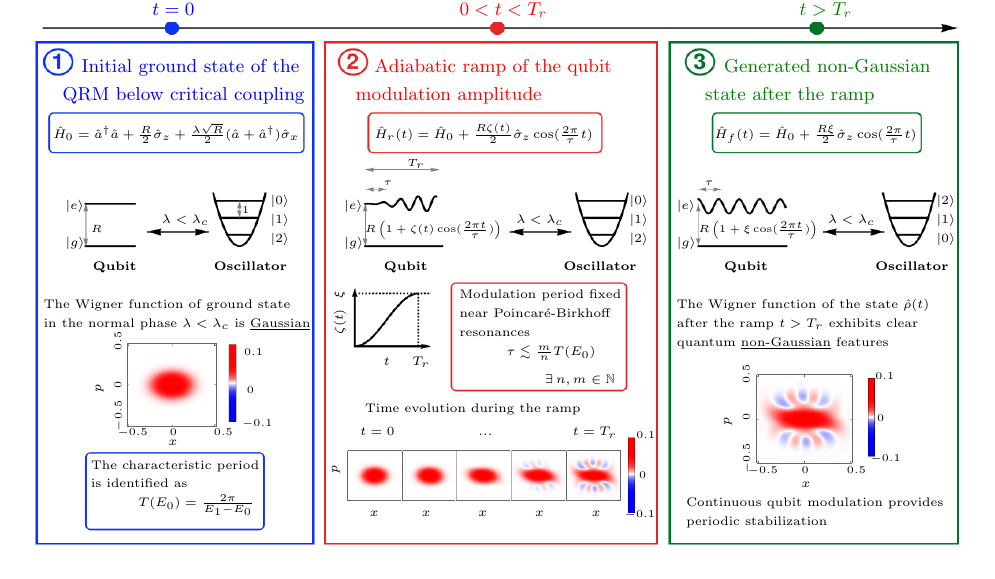}
    \caption{\label{protocol}
    Ramp protocol for the preparation of quantum non-Gaussian states
    in the QRM with coupling strength below the critical point
    $\lambda<\lambda_c$ and a weak qubit modulation.
    The protocol is divided into three basic steps.
    (1) $t=0$: preparation of the initial Gaussian state
    $|E_0\rangle$ as the ground state of the QRM in the normal
    phase $\lambda<\lambda_c$.
    The initial Hamiltonian has characteristic periods
    $T(E_k)=\frac{2\pi}{E_{k+1}-E_k}$,
    from which we identify the period corresponding to the
    ground state $T(E_0)$.
    (2) $0<t\leq T_r$: adiabatic ramp $\zeta(t)$ of the
    qubit-frequency modulation amplitude from $\zeta(0)=0$
    to $\zeta(T_r)=\xi$, with a fixed modulation period near
    Poincaré-Birkhoff resonances
    $\tau\lesssim\frac{m}{n}T(E_0)$,
    where $n,m$ are integers.
    (3) $t>T_r$: continuous qubit modulation with amplitude
    $\xi$ for periodic stabilization of the quantum
    non-Gaussian state after the ramp.
    }
\end{figure*}

\section{Introduction}

Time-periodic quantum systems and their associated quasistationary Floquet states \cite{Timberlake,segura2025quantum,wisniacki2015quantum,billam2009quantum,PhysRevE.109.014209,CauxPRL2018,PhysRevB.105.174205} have attracted considerable attention because periodic driving enriches the variety of accessible quantum phenomena while providing additional routes for coherent quantum control. Importantly, the high degree of tunability required for periodic modulation is experimentally available in a broad range of quantum platforms, including trapped-ion systems \cite{RMP2003,lv2017reconstruction,lv2018quantum} and superconducting circuit-QED architectures \cite{li2013motional,Solano2019}.

In the context of spin-boson systems, periodic modulation of spin frequencies has been proposed for effectively enhancing the coupling strength \cite{liu2024QRMdrive,Huang2016,Zhengemergentcat}. Beyond enhancing the interaction strength, fine-tuning of weak periodic modulation can induce resonant phenomena, leading to Floquet states with phase-space structures remarkably different from those obtained without modulation \cite{Timberlake,segura2025quantum,billam2009quantum,PhysRevE.109.014209}. This opens new possibilities for generating states with quantum non-Gaussian features that remain quasistationary under continuous modulation. Consequently, the design of protocols for the preparation and stabilization of Floquet states is an active topic of current research \cite{Desbuquois,bukov2015universal,3dpp-p2pr,nguyen2024programmable}.

Motivated by the rich structure of Floquet states in periodically driven systems \cite{billam2009quantum,PhysRevE.109.014209,wisniacki,wisniacki2015quantum,segura2025quantum,CauxPRL2018,PhysRevB.105.174205}, we investigate whether only weak spin-frequency modulation can be exploited to generate complex quantum non-Gaussian states at spin-boson coupling strengths far below the critical point. We consider the Quantum Rabi Model (QRM) \cite{rabi1936process} as a suitable model for this theoretical proposal due to the possibility of immediate experimental tests \cite{forn2010observation,braumuller2017analog,yoshihara2017superconducting,langford2017experimentally,crespi2012photonic,todorov2009strong,gunter2009sub,lv2018quantum,cai2021observation}. In the absence of modulation, the QRM exhibits a quantum phase transition separating the normal and superradiant phases \cite{PhysRevLett.115.180404}. Above the critical coupling, the ground state in the superradiant phase develops pronounced quantum non-Gaussian features \cite{catrabi}. However, the superradiant ground state must be reached by ramping the coupling across the critical point~\cite{cai2021observation,Zhengemergentcat},
and its quantum non-Gaussian features can be fragile against bosonic loss (see Appendix \ref{gramp}).

Here, we propose and analyze a ramp protocol that relies on weak periodic qubit modulation for generating persistent
quantum non-Gaussian states, rather than on strong coupling beyond the critical point. The protocol starts from the ground state of the QRM in the normal phase, with a coupling strength far below the critical point. While the coupling strength remains fixed, the qubit-frequency modulation is turned on adiabatically, such that the evolving state approximately follows the instantaneous Floquet state. We find that, when the modulation period is tuned close to Poincaré–Birkhoff resonant conditions, complex and robust quantum non-Gaussian features suddenly emerge. We further analyze the quantum non-Gaussian features of the generated state under realistic loss.

\section{Adiabatic Ramp Protocol}

Adiabatic ramp protocols~\cite{Desbuquois,Boyers2019,bukov2015universal,poletti2011slow} are particularly attractive for spin--boson systems because they are readily implementable with trapped-ions \cite{RMP2003,lv2017reconstruction,wunderlich2007robust,fabrikant2024cooling} and superconducting circuits \cite{kumar2016stimulated,PhysRevA.95.042330,3dpp-p2pr}. In the context of periodically driven systems, adiabatic ramps of the modulation amplitude have been proposed for the preparation of Floquet states~\cite{bukov2015universal,Desbuquois,goldman2014periodically,Berry_2009,3dpp-p2pr}. Here, we propose an adiabatic ramp of the qubit-modulation amplitude for generating \textcolor{orange!70!black}{persistent}
quantum non-Gaussian effects with coupling strengths below the critical point.

At the beginning of the ramp $t=0$, the Hamiltonian is the standard QRM  \cite{rabi1936process,Rossatto,xie2017quantum,braak2016semi,lv2018quantum} without modulation
\begin{equation}
\label{HQRM}
 \hat{H}_0=\hat{a}^\dagger\hat{a} + \frac{R}{2}\hat{\sigma}_z + \frac{\lambda\sqrt{R}}{2}(\hat{a}+\hat{a}^\dagger)\hat{\sigma}_x\,,
\end{equation}
where $\hat{a}^\dagger$ and $\hat{a}$ are the creation and annihilation operators of the bosonic mode, respectively, and $\vec{\sigma}=(\hat{\sigma}_x,\hat{\sigma}_y,\hat{\sigma}_z)=\left(|g\rangle\langle e|+|e\rangle\langle g|,\,i|e\rangle\langle g|-i|g\rangle\langle e|,\,|e\rangle\langle e|-|g\rangle\langle g|\right)$ are the Pauli matrices corresponding to the qubit. $R$ denotes the qubit frequency and serves as an effective system-size parameter, with the quantum phase transition emerging in the limit $R \to \infty$ \cite{PhysRevLett.115.180404}, while $\lambda$ is the coupling strength. The parity operator $\hat{\Pi} = (-1)^{\hat{n}} \hat{\sigma}_z$, with $\hat{n} = \hat{a}^\dagger \hat{a}$, is a conserved quantity. The critical coupling strength $\lambda_c=1$ separates the normal and superradiant phases of the QRM \cite{catrabi,PhysRevLett.115.180404} (See Appendix \ref{periodsapp} for a detailed analysis). The protocol initiates with the ground state of the QRM $|E_0\rangle$ as the initial state $\hat{\rho}(0)=|E_0\rangle\langle E_0|$. A detailed discussion of the preparation of the QRM ground state is presented in Appendix~\ref{gramp}. As shown there, for coupling strengths below the critical point $\lambda<\lambda_c$, the ground state can be prepared over relatively short timescales but remains Gaussian. Above the critical point $\lambda>\lambda_c$, the ground state develops quantum non-Gaussian features, but its preparation requires crossing the critical point
and its quantum non-Gaussian features are fragile against loss. For practical reasons, the protocol starts at a coupling strength below the critical point $\lambda<\lambda_c$, which is kept fixed throughout the protocol. Thus, the quantum non-Gaussian features generated subsequently are generated exclusively from the resonant qubit modulation.

At the endpoint of the ramp $t=T_r$, the Hamiltonian is that of the QRM under continuous qubit modulation \cite{liu2024QRMdrive,Huang2016,Zhengemergentcat},
\begin{eqnarray}
\label{HdrivenQRM}
\hat{H}_f(t)&=&\hat{H}_0 + \frac{R\xi}{2} \Vec{\eta} \cdot \hat{\Vec{\sigma}} \cos(\frac{2\pi}{\tau} t)\,,
\end{eqnarray}
where $\xi$ and $\tau$ are the amplitude and the modulation period, respectively.  The unit vector \(\Vec{\eta}\) specifies the direction of the qubit modulation. Experimentally, two modulation directions, \(\Vec{\eta} = \Vec{z}\) and \(\Vec{\eta} = \Vec{x}\), are straightforward in both circuit-QED platforms~\cite{johansson2006vacuum,li2013motional,Solano2019} and trapped-ion systems~\cite{RMP2003,lv2017reconstruction,lv2018quantum}.
When the modulation is applied along the direction $\Vec{\eta} = \Vec{z}$, the parity operator $\hat{\Pi} = (-1)^{\hat{n}} \hat{\sigma}_z$ holds as a conserved quantity. When the modulation is applied along the direction $\Vec{\eta} = \Vec{x}$, the parity is broken. The Floquet state $|F_0\rangle$ of the time-periodic Hamiltonian~\eqref{HdrivenQRM}, selected as the state minimizing $\langle F_k|\hat{H}_0|F_k\rangle$, is adiabatically connected to the QRM ground state $|E_0\rangle$ sufficiently far from level crossings. This state is the ideal target state  $\hat{\rho}_{\mathrm{tar}}=|F_0\rangle\langle F_0|$ of the protocol. As Floquet states are quasistationary under periodic unitary evolution~\cite{bukov2015universal}, the continuous qubit modulation is expected to provide temporal stability to the prepared state after the ramp.

The modulation period $\tau$ is selected following the characteristic periods of the QRM, defined from the spectral spacing of the initial Hamiltonian~\eqref{HQRM} as $T(E_k)=2\pi/(E_{k+1}-E_k)$~\cite{NADER2023PLA}. In the semiclassical limit $R\to\infty$, these characteristic periods approach the corresponding classical orbital periods at energy $E_k$ (see Appendix~\ref{periodsapp} for further details). According to the Poincar\'e--Birkhoff theorem~\cite{birkhoff1926extension}, a weak time-periodic perturbation reorganizes the classical phase-space structure near resonances satisfying $\tau \approx \frac{m}{n}T(E_k)$, with $m,n\in\mathbb{N}$, through the destruction of the corresponding invariant tori and the formation of resonant structures (see Appendix~\ref{resonant} for a detailed discussion). A related restructuring is also reflected in the phase-space structure of the Floquet states near these resonant conditions, with the resulting structures exhibiting system-size scaling~\cite{segura2025quantum,billam2009quantum}.
In particular, the Floquet state $|F_0\rangle$ develops complex quantum non-Gaussian structures in the vicinity of the Poincar\'e--Birkhoff resonances (see Appendix~\ref{Floquetapp}), even for weak modulation amplitudes $\xi\ll1$, and coupling strengths below the critical point $\lambda<\lambda_c$.
We therefore deliberately select the modulation period $\tau$
close to the resonant conditions associated with the ground-state
characteristic period, $\tau \lesssim \frac{m}{n}T(E_0)$, while a large qubit frequency $R$ favors the emergence of well-resolved resonant structures in the Floquet states.

Having identified the desired Floquet state $|F_0\rangle$ at the endpoint of the ramp, we propose the following time-dependent Hamiltonian, which is valid during and after the ramp
\begin{equation}
\label{Hramp}
\hat{H}_{r} (t) = \hat{H}_{0} + \frac{R\zeta(t)}{2} \Vec{\eta} \cdot \hat{\Vec{\sigma}} \cos(\frac{2\pi}{\tau} t)\,,
\end{equation}
where $\zeta(t)$ is the smooth cosine ramp \cite{Flaschner,nguyen2024programmable} programmable in experiments
\[
\label{ramp}
\zeta(t) =
\begin{cases}
\frac{\xi}{2}\left(1-\cos\left(\frac{\pi}{T_r}t\right)\right), & 0 \leq t \leq T_r \\
\xi, & t > T_r
\end{cases}
\]
The time-dependent Hamiltonian (\ref{Hramp}) connects the QRM Hamiltonian (\ref{HQRM}) at the beginning of the ramp $\zeta(0)=0$ with the Hamiltonian of the QRM under continuous modulation (\ref{HdrivenQRM}) at the end of the ramp $\zeta(T_r)=\xi$. During the whole protocol, the modulation period $\tau$ and the
coupling strength $\lambda$ remain fixed. According to the generalized adiabatic theorem for time-periodic systems \cite{dodin2021generalized}, if the modulation amplitude increases sufficiently slowly, the state approximately follows the corresponding instantaneous Floquet state $|F_0(t)\rangle$. The modulation amplitude is increased adiabatically, rather than using abrupt quenches \cite{d2014long,nag2019out}, to favor the temporal stability of the Floquet states over rapidly changing quantum non-Gaussian features.

Although longer ramp durations improve adiabaticity, shorter ramps are preferred in practice to avoid decoherence effects that destroy the quantum non-Gaussian features. To take into account decoherence effects during the adiabatic ramp, we numerically solve the Lindblad master equation \cite{breuer2002theory} including loss
\begin{equation}
\label{Lindblad}
\frac{d\hat{\rho}(t)}{dt}
=
-i[\hat{H}_{r}(t),\hat{\rho}]
+
\kappa\, D[\hat{a}]\,\hat{\rho}
\end{equation}
where the superoperator
$
D[\hat{a}]\,\hat{\rho}
=
\hat{a}\hat{\rho}\hat{a}^\dagger
-
\frac{1}{2}
\left(
\hat{a}^\dagger \hat{a}\hat{\rho}
+
\hat{\rho}\hat{a}^\dagger \hat{a}
\right)
$
is the standard Lindblad dissipator. The initial condition for the Lindblad master equation (\ref{Lindblad}) is $\hat{\rho}(0) = |E_0\rangle \langle E_0|$. At the end of the ramp $t=T_r$, the state is expected to approach the target Floquet state $\hat{\rho}_{\mathrm{tar}}=|F_0\rangle\langle F_0|$. However, the central goal of the protocol is not to maximize the fidelity with this state, but to generate complex quantum non-Gaussian features during the ramp.  Therefore, we focus dominantly on their evaluation in a directly measurable Wigner function.
The protocol is divided into three basic steps and summarized in Fig.~\ref{protocol}.

\section{Fidelity and phase space quantifiers}

To quantify the similarity with respect to the ideal target state, we use i) the Uhlmann fidelity
\begin{equation}
\label{fidelity}
\mathcal{F}(\hat{\rho}(t),\hat{\rho}_{tar})=
\left(\mathrm{Tr}
\sqrt{\sqrt{\hat{\rho}_{tar}}
\hat{\rho}(t)
\sqrt{\hat{\rho}_{tar}}}\right)^2
\end{equation}
between the evolved state $\hat{\rho}(t)$ and the target state $\hat{\rho}_{tar}=|F_0\rangle\langle F_0|$. If the two states are pure, for example at the beginning of the ramp $t=0$, the fidelity reduces to the overlap between the ground state of the QRM and the target Floquet state $|\langle E_0|F_0\rangle|^2$. Fidelity, as a global measure of the similarity between the generated and target states, may be insensitive to fine structures in phase space.

As another fidelity-based measure, we consider ii) the fidelity susceptibility \cite{PhysRevE.99.050102,braun2018quantum}, which quantifies the sensitivity of a quantum state to variations in an unknown parameter $\phi$.  The fidelity susceptibility is given by
\begin{equation}
\label{QFIfid}
F_s[\hat{\rho}_b(t),\hat{A}] =
\lim_{\phi\to0}
\frac{
1-\mathcal{F}\left(
\hat{\rho}_b(t),
\hat{\rho}_b(t,\phi)
\right)
}{\phi^2},
\end{equation}
where \(\hat{\rho}_b(t)= {\rm Tr}_q[\hat{\rho}(t)]\) denotes the bosonic density matrix and
$
\hat{\rho}_b(t,\phi)
=
e^{-i\phi\hat{A}}
\hat{\rho}_b(t)
e^{i\phi\hat{A}}
$ is the encoded state. For a given parameter encoding, the fidelity susceptibility is proportional \cite{LIU2014167} to the quantum Fisher information \cite{geeza2014quantum,Paris2009}. Of particular interest are phase transformations generated by the bosonic number operator, $\hat{A}=\hat{n}=\hat{a}^{\dagger}\hat{a}$, for which the fidelity susceptibility quantifies the sensitivity to phase space rotations \cite{Fadel_2025}. As a reference, we subtract the fidelity susceptibility of the initial state and normalize by the susceptibility of coherent states $\bar{F}_s[\hat{\rho}_b(t),\hat{n}]=(F_s[\hat{\rho}_b(t),\hat{n}]-F_s[\hat{\rho}_b(0),\hat{n}])/\bar{n}$ with $\bar{n}={\rm Tr}\left(\hat{\rho}_b(t)\hat{n}\right)$. During the protocol, we compare the susceptibility of the generated state with that of the reference Gaussian state with identical covariance matrix \cite{Genoni2007}.

Beyond the parallel analysis of fidelity and its susceptibility, the primary goal of the protocol is to generate quantum non-Gaussian effects in phase space at coupling strengths below the critical point. To witness the emergence of such effects, we use the experimentally measurable bosonic Wigner function \cite{hillery1984distribution}
\begin{equation}
\label{Wignerqpt}
W(x,p,t)=\frac{R}{\pi}\int_{-\infty}^{\infty} dy e^{-2iypR} \langle x+y|\hat{\rho}_b(t)| x-y\rangle\,,
\end{equation}
where $\hat{\rho}_b (t)= {\rm Tr}_q (\hat{\rho}(t))$ is the bosonic density matrix. The bosonic Wigner function can be directly sampled in trapped-ion and superconducting-circuit platforms, without requiring full density-matrix tomography, by measuring the displaced parity at selected phase-space points~\cite{bertet2002direct,wang2009decoherence,wang2016schrodinger}.
To quantify quantum non-Gaussian effects during the ramp, we use iii) the Wigner negativity volume
\begin{equation}
\label{negatitity}
  \delta(t)=\int\int (|W(x,p,t)|-W(x,p,t))  dx dp\,,
\end{equation}
which is widely considered a standard directly measurable quantifier of quantum non-Gaussian aspects in phase space \cite{AnatoleKenfack_2004}.

As a measure of complexity in phase space, we introduce a iv) negativity-resolved extension of the root-mean-square (RMS) Fourier radius of the Wigner function~\cite{chi},
\begin{equation}
\chi_-^2(t)=
\frac{\displaystyle \int dx\,dp\, |\nabla W(x,p,t)|^2\,\Theta[-W(x,p,t)]}
{\displaystyle 4\pi^2\int dx\,dp\, W^2(x,p,t)}\,,
\label{chi}
\end{equation}
where $\nabla=\vec{x}\partial_x+\vec{p}\partial_p$ is the phase space gradient. The conventional RMS Fourier radius
quantifies the fine-scale complexity of the Wigner function in phase space. By introducing the Heaviside function $\Theta[-W(x,p,t)]$, we restrict the numerator to the negative regions of the Wigner function while retaining the global normalization. Consequently, $\chi_-^2$ quantifies the fine-scale complexity of the negative Wigner function.

Finally, for comparison of the negative regions of the Wigner function with respect to the target state, we propose the v) negativity overlap
\begin{widetext}
\begin{equation}
\label{autocor}
N(t)=
\begin{cases}
\displaystyle
\frac{
\int dx\,dp\,
W_{tar}(x,p)W(x,p,t)
\Theta[-W_{tar}(x,p)]
\Theta[-W(x,p,t)]
}{
\sqrt{\int dx\,dp\,W_{tar}^2(x,p)\Theta[-W_{tar}(x,p)]}
\sqrt{\int dx\,dp\,W^2(x,p,t)\Theta[-W(x,p,t)]}
},
& \delta_{tar}>0 \ \text{and}\ \delta(t)>0,\\[2ex]
0,
& \delta_{tar}=0 \ \text{or}\ \delta(t)=0.
\end{cases}
\end{equation}
\end{widetext}
where $W_{tar}(x,p)$ is the Wigner function of the target Floquet state, $W(x,p,t)$ is the Wigner function of the state during the ramp, and $\Theta(x)$ is the Heaviside function. The negativity overlap quantifies similarity between the negative part of the Wigner function of the evolving state and the target Floquet state. The denominator is introduced to normalize the overlap such that $N(t)=1$ for identical negative Wigner functions, whereas $N(t)=0$ indicates either that one of the Wigner functions is positive or that their negativities occupy disjoint regions of phase space.

\section{Results}

\begin{figure*}[!htbp]
    \centering
    \includegraphics[width=\textwidth]{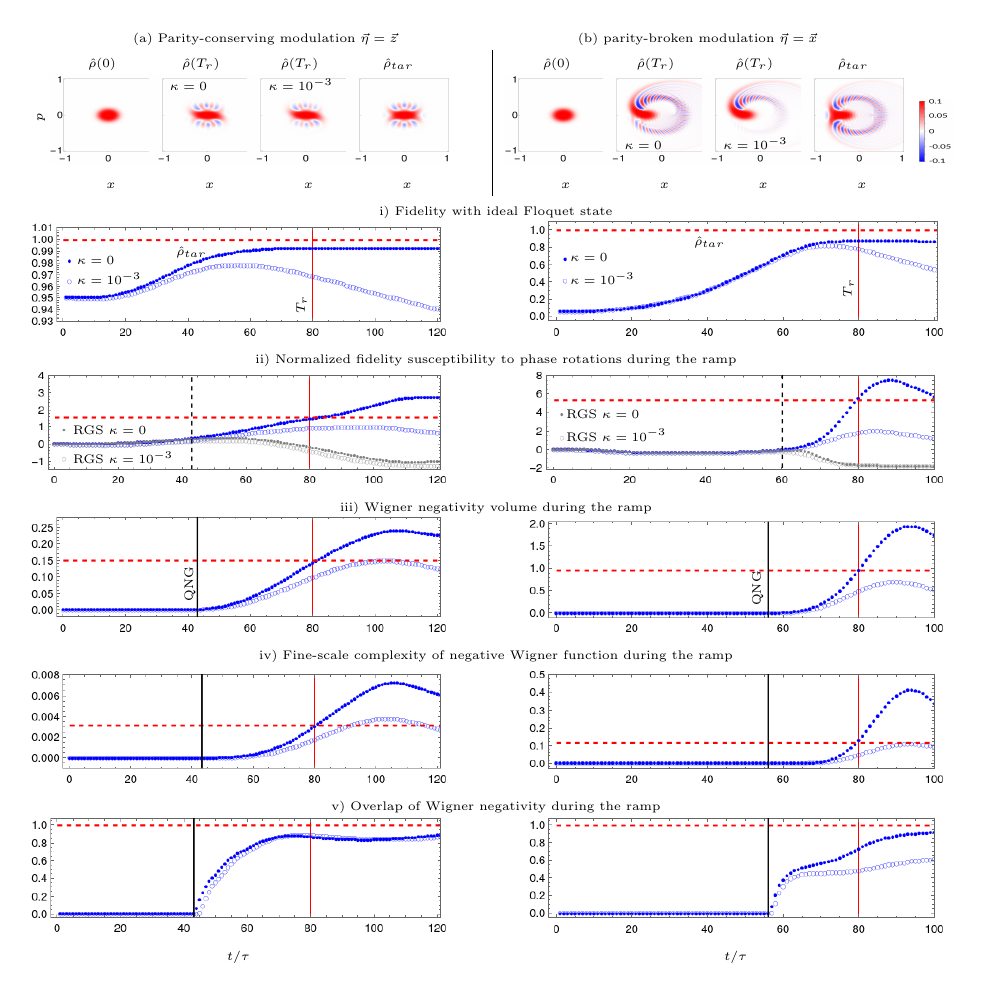}
    \caption{\label{example_protocol} Ramp protocol to approach the Floquet quantum non-Gaussian states in the QRM below the critical point.
 (a) parity-conserving  qubit modulation $\Vec{\eta} = \Vec{z}$ with resonant period $\tau = 4.29 \lesssim \tfrac{1}{2}T(E_0)$.
 (b) parity-broken  qubit modulation $\Vec{\eta} = \Vec{x}$ with resonant period $\tau = 8.2 \lesssim T(E_0)$.
    In the first row, we show the Wigner functions of the initial state $\hat{\rho}(0)$, the state at the end of the ramp $\hat{\rho}(T_{r})$ with and without loss $\kappa$, and the target state $\hat{\rho}_{tar}=|F_0\rangle\langle F_0|$. i) Fidelity with respect to the target Floquet state ${\mathcal F}(\hat{\rho}(t),\hat{\rho}_{tar})$. ii) Normalized fidelity susceptibility to phase rotations $\bar{F}_s[\hat{\rho}_b(t),\hat{n}]$. iii) Wigner negativity volume $\delta(t)$. iv) Fine-scale complexity of negative Wigner function $\chi^2_-(t)$. v) Overlap of Wigner negativity $N(t)$ with respect to the target Floquet state. All quantifiers i-v) are presented as a function of normalized time $t/\tau$. In all panels, filled circles denote the evolution without loss $\kappa=0$, whereas empty circles correspond to the evolution with loss $\kappa=10^{-3}$. The red horizontal dashed line corresponds to the target Floquet state $\hat{\rho}_{tar}$,  the vertical red line marks the end of the ramp at $T_{r}=80\tau$ and the vertical black line marks the onset of the quantum non-Gaussian (QNG) effects.  Color gray corresponds to the reference Gaussian state (RGS) with the same covariance matrix, while the dashed vertical line indicates the deviation of the generated state from this reference Gaussian state. The initial state is the ground state of the QRM, $\hat{\rho}(0)=|E_0\rangle \langle E_0|$, prepared in the normal phase with coupling strength below the critical point $\lambda = \tfrac{1}{\sqrt{2}} < \lambda_c$ and qubit frequency $R = 50$. The characteristic period for this configuration of parameters is $T(E_0)=8.8$. The final modulation amplitude is $\xi = \frac{1}{20}$.
    }
\end{figure*}

To demonstrate the efficacy of the protocol for generating quantum non-Gaussian effects at coupling strength far below the critical point, we consider two representative cases: (a) parity-conserving qubit modulation $\Vec{\eta}=\Vec{z}$, with a modulation period near resonance $\tau=4.29 \lesssim \frac{1}{2}T(E_0)$, and (b) parity-broken qubit modulation $\Vec{\eta}=\Vec{x}$, with modulation period near resonance $\tau=8.2 \lesssim T(E_0)$. In both cases, the system is initially prepared in the ground state of the QRM within the normal phase $\lambda=\frac{1}{\sqrt{2}}<\lambda_c$ far below the critical point and a finite qubit frequency $R=50$. For this configuration of parameters, the characteristic period corresponding to the ground state is $T(E_0)=8.8$. The modulation amplitude is then slowly ramped up to the weak final value $\xi=1/20$. The ramp duration is chosen to be $80$ modulation cycles, i.e., $T_r=80\tau$, limited by the available computational resources. The modulation periods $\tau$ are deliberately chosen close to the corresponding resonant conditions yet sufficiently far from level crossings between Floquet branches to preserve the adiabatic continuation between the QRM ground state $|E_0\rangle$ and the target Floquet state $|F_0\rangle$. The results are shown in Fig. \ref{example_protocol} including the Wigner function of the initial $\hat{\rho}(t=0)=|E_0\rangle\langle E_0|$, final $\hat{\rho}(t=T_r)$ and target states $\hat{\rho}_{tar}=|F_0\rangle\langle F_0|$. Throughout the ramp protocol we follow the fidelity i), phase sensitivity ii) and detailed quantifiers of quantum non-Gaussian effects in phase space iii)-v).

For the parity-conserving  modulation (a), the generated state has a high fidelity with the target state, increasing from $\sim0.95$ to $\sim0.99$ at the end of the ramp. However, the fidelity is sensitive to loss. For the parity-broken  modulation (b), the initial state has a very low fidelity with the target state $\sim0.05$. Despite the greater complexity of the target state, the lossless fidelity increases substantially during the ramp, reaching approximately $\sim0.85$ at the end of the ramp. In both cases, the fidelity remains approximately constant after the ramp. With bosonic loss, the fidelity decreases gradually after the ramp. This indicates that the temporal stability of the generated states is affected by the loss.

The fidelity susceptibility to phase rotations provides further insight into the generated states beyond their fidelity with the target state. For parity-conserving modulation (a), the normalized fidelity susceptibility grows monotonically together with that of its reference Gaussian state, and after $43$ modulation cycles, they deviate. Interestingly, the normalized fidelity susceptibility of the generated state keeps increasing while that of its reference Gaussian state decreases having the same covariance matrix. At the end of the ramp, the generated state essentially reaches the same susceptibility as the ideal target state, while the reference Gaussian state falls below the initial normalized fidelity susceptibility ( therefore, the normalized fidelity susceptibility reaches negative values). This effect already hints that the quantum non-Gaussian features of the generated state enhance the normalized fidelity susceptibility far beyond that of its reference Gaussian state with the same covariance matrix. This enhanced susceptibility remains robust against loss and continues to grow after the ramp under continuous modulation. For parity-broken modulation (b), the normalized fidelity susceptibility of the generated state deviates from that of its reference Gaussian state after $59$ modulation cycles. After that, the normalized fidelity susceptibility of the generated state is drastically enhanced and essentially reaches the same value as the target Floquet state. The enhanced normalized fidelity susceptibility of the generated state is more sensitive to loss. Without loss, it overshoots the target value after the ramp and then decreases, for the same reason as the negativity discussed below.

For a more detailed analysis of the generated state we proceed to examine the detailed quantifiers of quantum non-Gaussian effects. For the case of parity-conserving  modulation (a), quantum non-Gaussian effects appear after $43$ modulation periods, exactly when the normalized fidelity susceptibility to phase rotations deviates from that of its reference Gaussian state. For the case of parity-broken  modulation (b), quantum non-Gaussian effects appear after $57$ modulation periods, shortly before the normalized fidelity susceptibility to phase rotations deviates from that of its reference Gaussian state. The negativity volume and the negative-resolved RMS Fourier radius confirm that the state generated with the parity-broken  modulation in (b) contains significantly more negativity and is more complex in phase space than the state generated with the parity-conserving  modulation (a). This corresponds to the higher susceptibility to phase rotations of the state generated with the parity-broken  modulation (b). In the case of parity-conserving  modulation (a), the negativity volume at the end of the ramp reaches essentially the same value as that of the target Floquet state $\sim0.15$, with a complexity of the negative Wigner function $\sim3\times10^{-3}$.
The state generated with the parity-broken  modulation (b) reaches a Wigner negativity volume close to unity $\sim0.96$, and a complexity of the negative Wigner function $\sim0.12$, values similar to those of the target Floquet state.

After the ramp, the fidelity remains constant for $\kappa=0$, whereas the Wigner negativity volume and the complexity of the negative Wigner function first continue to grow under continuous modulation, transiently exceeding the values of the target Floquet state [by a factor of about 1.5 in (a) and 2 in (b)], and then decrease. This behavior has a simple origin. For $t>T_r$ and vanishing dissipation, the stroboscopic evolution is generated by the Floquet operator, which conserves the populations of the Floquet states; therefore, the fidelity with $|F_{0}\rangle$ remains constant after the ramp. Any further change of the phase-space quantifiers originates from coherences between $|F_{0}\rangle$ and other Floquet states admixed during the finite ramp. The quantifiers are thus quasi-periodic rather than stationary. Importantly, in the presence of loss, the negativity remains substantial throughout the simulated time window, in contrast with the superradiant ground state (see Appendix~\ref{gramp}).

At the end of the ramp, the negativity overlap reaches $\approx 0.8$, although the negativity volume and the fine-scale complexity of the negative Wigner function are close to those of the target state. This is because some negative regions may be slightly displaced relative to one another in phase space. After the ramp, the overlap remains approximately constant in case (a) and increases further in case (b). Hence, while the magnitude of the negativity varies, the negative regions remain located where the target Floquet state predicts them.

 After analyzing all the quantifiers, we conclude that a comparison with a preselected complex target state provides only a limited picture of the quantum non-Gaussian effects generated during the ramp protocol. Instead, the combined analysis of the negativity volume and fine-scale complexity provides a more reliable quantification of the protocol efficacy in generating complex quantum non-Gaussian states using a minimal toolbox. The quantum non-Gaussian features of the generated state remain robust in the presence of loss, indicating that they are achievable and testable. This robustness against loss can be contrasted with that of the QRM ground state prepared by slowly increasing the coupling strength beyond the critical point (see Appendix \ref{gramp}). In the latter case, the loss significantly degrades the quantum non-Gaussian features of the QRM ground state, as indicated by the strong reduction in both the negativity volume and the fine-scale complexity of the negative Wigner function.


\section{Conclusions}

The results presented in this work demonstrate that complex and robust quantum non-Gaussian features can be induced by a weak time-periodic qubit modulation, even for coupling strengths far below the critical point. Remarkably, the simple ramp protocol enables the generation of substantial Wigner negativity and highly complex negative structures of the Wigner function extending over broad regions of phase space, which we investigate using detailed quantifiers of quantum non-Gaussian effects. These essential aspects are not immediately visible in the analysis of fidelity and its susceptibility. It therefore promotes quantum non-Gaussian quantifiers for further investigation of the critical effects in time-periodic systems. The protocol is experimentally implementable with current platforms and therefore provides an attractive route for the generation and investigation of complex quantum non-Gaussian effects in phase space below the critical points.

\section{Acknowledgments}
D.J.N. and R. F. acknowledge the support from grant CZ.02.01.01/00/22\_008/0004649
(QUEENTEC) provided by the EU and MEYS of the Czech Republic.  D.J.N. also acknowledges QuantERA project Q3ML (9C26002) of the MEYS of the Czech Republic. R.F. was also supported by QuantERA project CLUSSTAR (8C2024003) of the MEYS of the Czech Republic. Projects
CLUSSTAR and Q3ML have received funding from the EU Horizon Programme under Grant Agreement No. 731473 and 101017733
(QuantERA). R.F. also acknowledges the EU Horizon Programme
under Grant Agreement No.101080173 (CLUSTEC). P. C.  and P. S.  acknowledge grant 25-16056S of the Czech Science Foundation. J.C -C, has received a fellowship (Grant No. 493338) from Secretaría de Ciencias y Humanidades, Tecnología e Innovación (SECIHTI).

\section{Data Availability}

All data and code needed to reproduce the results in this paper are available online at  \url{https://github.com/djuliannader/RabiModel}

The Lindblad master equation \eqref{Lindblad} was solved numerically using the Julia package \textit{DifferentialEquations.jl}~\cite{DiferentialEquations}. The initial density matrices were constructed using the product of a truncated Fock basis and the qubit degrees of freedom.  In all calculations, the Fock basis was truncated at $N=50$. The Wigner functions \eqref{Wignerqpt} were computed using the \textit{QuantumOptics.jl} framework~\cite{QuantumOptics}, and the integrals required to evaluate the negative volume \eqref{negatitity} were calculated using the Julia package \textit{HCubature}~\cite{HCubature}. The number of time subintervals in the Trotter-Suzuki decomposition of the Floquet operator was fixed to $l=1000$. The Poincar\'e maps were obtained by evaluating the classical equations of motion with the semiclassical time-periodic Hamiltonian \eqref{Hsemiclassicalperiodic} using the \textit{Mathematica} package \textit{NDSolve}~\cite{MathematicaNDSolve}.

\bibliography{Bibliography}

\appendix

\section{Semiclassical limit and characteristic periods}\label{periodsapp}

\begin{figure}[t!]
    \centering
    \includegraphics[width=\columnwidth]{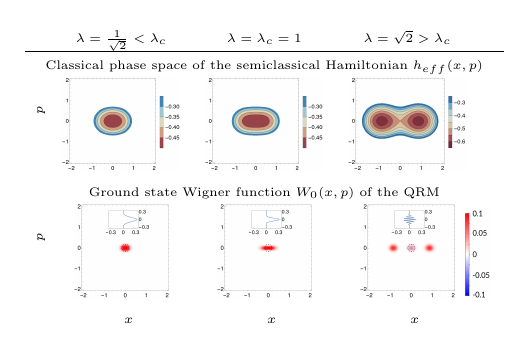}
    \caption{\label{wignerns} 
   (Top row)  Classical phase space corresponding to the semiclassical Hamiltonian (\ref{Hsemiclassical}) of the QRM. The color scheme represents a set of trajectories with neighboring energies. (Bottom row) Ground state Wigner function of the QRM (\ref{HQRM}). The dotted circle represents the ground state of the free oscillator Hamiltonian obtained in the limit of vanishing coupling $\lambda\to0$. The inset in each Wigner function contains the cut along the momentum axis. The coupling is chosen at specific values smaller and larger than the critical $\lambda_c=1$. The qubit frequency is fixed $R=50$.
    }
\end{figure}

For a semiclassical approximation to the QRM, where the spectrum becomes effectively continuous and nonanalytic points associated with quantum phase transitions can emerge, one can use a mean-field approximation \cite{Stransky2021,catdeath2024,irish2014oscillator,Puebla2016}. First, we take the expectation value of the QRM Hamiltonian in the oscillator subspace, $\hat{H}_\alpha=\langle\alpha|\hat{H}_0|\alpha\rangle$, where $|\alpha\rangle=\hat{D}(\alpha)|0\rangle={\rm exp}(\alpha\hat{a}^\dagger-\alpha^*\hat{a})|0\rangle$ is a coherent state with displacement parameter $\alpha=\sqrt{\frac{R}{2}}(x+ip)$.
Notice that the coherent state becomes more localized as the qubit frequency $R$ increases.
The eigenvalues of the resulting expectation value correspond to the effective semiclassical Hamiltonians of the spin-up and spin-down subspaces. We can write these effective semiclassical Hamiltonians in terms of the real and imaginary parts of the displacement parameter as $x$ and $p$, obtaining
\begin{equation}
\label{Hsemiclassical}
h_{eff}^{\pm}=\frac{\langle H \rangle}{R} =\frac{p^2+x^2}{2}\pm \frac{1}{2}\sqrt{1 + 2\lambda^2x^2}\,.
\end{equation}
This effective Hamiltonian is composed of an oscillator term and a square-root term that accounts for the off-resonant interaction with the qubit. The QPT of the QRM can be understood as a qualitative change in the effective potential from a single-well structure for $\lambda<1$ (normal phase) to a double-well structure for $\lambda>1$ (superradiant phase), leading to a nonanalyticity in the ground-state energy at the critical point i.e. a quantum phase transition.  The qualitative change of the ground state across the critical point is well illustrated by the bosonic Wigner function (\ref{Wignerqpt}). As shown in Fig.~\ref{wignerns}, the ground state in the normal phase is Gaussian, i.e. positive everywhere in phase space, while in the superradiant phase it becomes non-Gaussian as the state separates into two components and quantum interference fringes emerge in the middle of phase space (see \cite{catrabi} for a more detailed discussion). The Wigner function is in good correspondence with the phase-space picture provided by the semiclassical Hamiltonian (\ref{Hsemiclassical}).

From the semiclassical perspective, the characteristic periods of the QRM correspond to the periods of the classical orbits of energy \eqref{Hsemiclassical}, and are given by \cite{NADER2023PLA}
\begin{equation}
\label{classicalT}
T(E)=\int_{x_1}^{x_2}\frac{\sqrt{2}}{\sqrt{E-V(x)}}dx\,,
\end{equation}
where $V(x)$ is the potential energy term in (\ref{Hsemiclassical}), $x_{1,2}$ are the classical turning points corresponding to the energy $E$. In the normal phase $\lambda < \lambda_c$, the potential contains a single well, and there are only two classical turning points for any energy $E$. In contrast, in the superradiant phase $\lambda>\lambda_c$, the potential develops degenerate wells with a central barrier whose maximum at $x=0$ is $E=-1/2$. For energies below the central barrier $E< -1/2$, there are four classical turning points, and for energies above the barrier $E>-1/2$, only two remain. Consequently, the period (\ref{classicalT}) exhibits a non-analyticity at the energy corresponding to the local maximum, $E_{ESQPT} = -1/2$. The nonanalyticity at the critical energy $E_{ESQPT}$ is the signature of an excited-state quantum phase transition (ESQPT) \cite{cejnar2021excited}.

In the quantum realm, the characteristic period can be found through its relation with the Density of States (DOS) \cite{NADER2023PLA}
\begin{equation}
\label{quantumT}
T(E_k)=2\pi\rho(E_k)
\end{equation}
where the DOS is given by
\begin{equation}
\rho(E_k)=
 \begin{cases}
  \frac{2}{E_{2(k+1)}-E_{2k}},
 \,\, \text{for $E_k<E_{ESQPT}$} \\
 \frac{1}{E_{k+1}-E_{k}}, \,\,  \text{for $E_k>E_{ESQPT}$}
 \end{cases}
\end{equation}
where $E_k$ are the eigenvalues of $\hat{H}_0$ and the discontinuity arises because the energy levels of different parity $\hat{\Pi}$ below $E_{ESQPT}$ are quasidegenerated. Indeed, the energy gap between consecutive levels of opposite parity becomes exponentially small as a function of the coupling strength $\lambda$ \cite{PhysRevLett.115.180404}.

\begin{figure}[t!]
    \centering
    \includegraphics[width=\columnwidth]{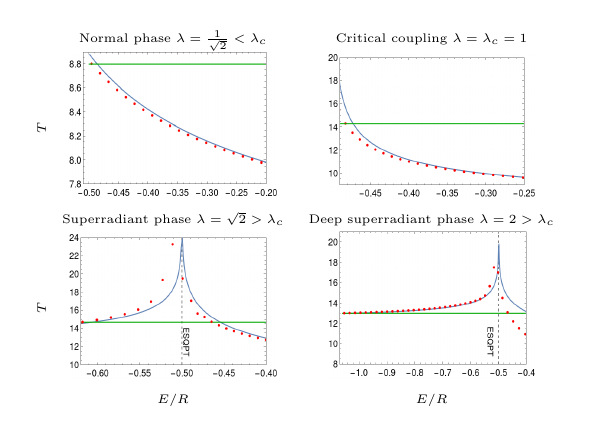}
    \caption{\label{periods}  Characteristic periods $T$ of the QRM without modulation as a function of the normalized energy $E/R$. The blue solid line corresponds to the classical periods from (\ref{classicalT}) and the red points to the quantum periods from (\ref{quantumT}). The dashed vertical line indicates the logarithmic divergence during the ESQPT. The coupling strength is selected at different values, both in the normal $\lambda<1$ and superradiant phase $\lambda>1$, while the qubit frequency is fixed $R=50$. The red dots approach the solid blue line as the qubit frequency $R$ increases. The green horizontal lines represent the characteristic quantum period $T(E_0)$ corresponding to the energy of the ground state.  An approximation to the resonant periods $\tau$ can be obtained analytically by expanding the effective potential $v_{eff}(x)$ of the semiclassical Hamiltonian (\ref{Hsemiclassical}) in powers of the position $x$  and identifying the harmonic frequency $T(E_0) \approx \frac{2\pi}{\sqrt{1-\lambda^2}}\,$.
    }
\end{figure}

The quantum (\ref{quantumT}) and classical (\ref{classicalT}) characteristic periods of the QRM are shown in Fig. \ref{periods} as a function of the normalized energy $E/R$. The results exhibit a relatively good correspondence for a finite qubit frequency $R=50$ and improve as $R$ increases \cite{Stransky2021,catdeath2024}. In the normal phase $\lambda<\lambda_c$, the period decreases smoothly as a function of energy. However, once the coupling exceeds the critical point $\lambda>\lambda_c$, both the period $T$ and the DOS, connected by (\ref{quantumT}), exhibit a discontinuity at the energy $E_{ESQPT}=-1/2$. This nonanalytic point is the signature of the ESQPT of the QRM \cite{Stransky2021,Puebla2016}. ESQPTs are typically accompanied by some interesting consequences in the dynamics and tunneling \cite{chavez2023spectral,Prado2023,Nader2021PRE}. Of particular importance for our study is the identification of the characteristic period associated with the ground state, $T(E_0)$, which we evaluate using (\ref{quantumT}).

\section{Ground-state preparation across the critical point}\label{gramp}

The ground state of the QRM without periodic modulation can be prepared from the initial uncoupled state $|\Psi_0\rangle=|0\rangle\otimes|g\rangle$, by adiabatically increasing the coupling strength $\lambda$ \cite{lv2018quantum,cai2021observation,pedernales2015quantum}. The corresponding time-dependent Hamiltonian is
\begin{equation}
\hat{H}(t)=\hat{a}^\dagger\hat{a}
+\frac{R}{2}\hat{\sigma}_z
+\lambda(t)\frac{\sqrt{R}}{2}
(\hat{a}^\dagger+\hat{a})\hat{\sigma}_x ,
\end{equation}
where the coupling increases according to the smooth ramp
\begin{equation}
\lambda(t)=\frac{\lambda_f}{2}\left[
1-\cos\left(\frac{\pi t}{T_r}\right)
\right],
\end{equation}
which connects the uncoupled initial configuration, $\lambda(0)=0$, to the final coupling strength $\lambda(T_r)=\lambda_f$ over the ramp duration $T_r$. If the ramp period is sufficiently large, the state approximately follows the instantaneous ground state of the QRM according to the adiabatic theorem \cite{born1928beweis,sakurai2020modern}. In the presence of loss, the evolution of the density matrix during the ramp is obtained by solving the master equation \eqref{Lindblad}.

\begin{figure*}[!htbp]
    \centering
    \includegraphics[width=\textwidth]{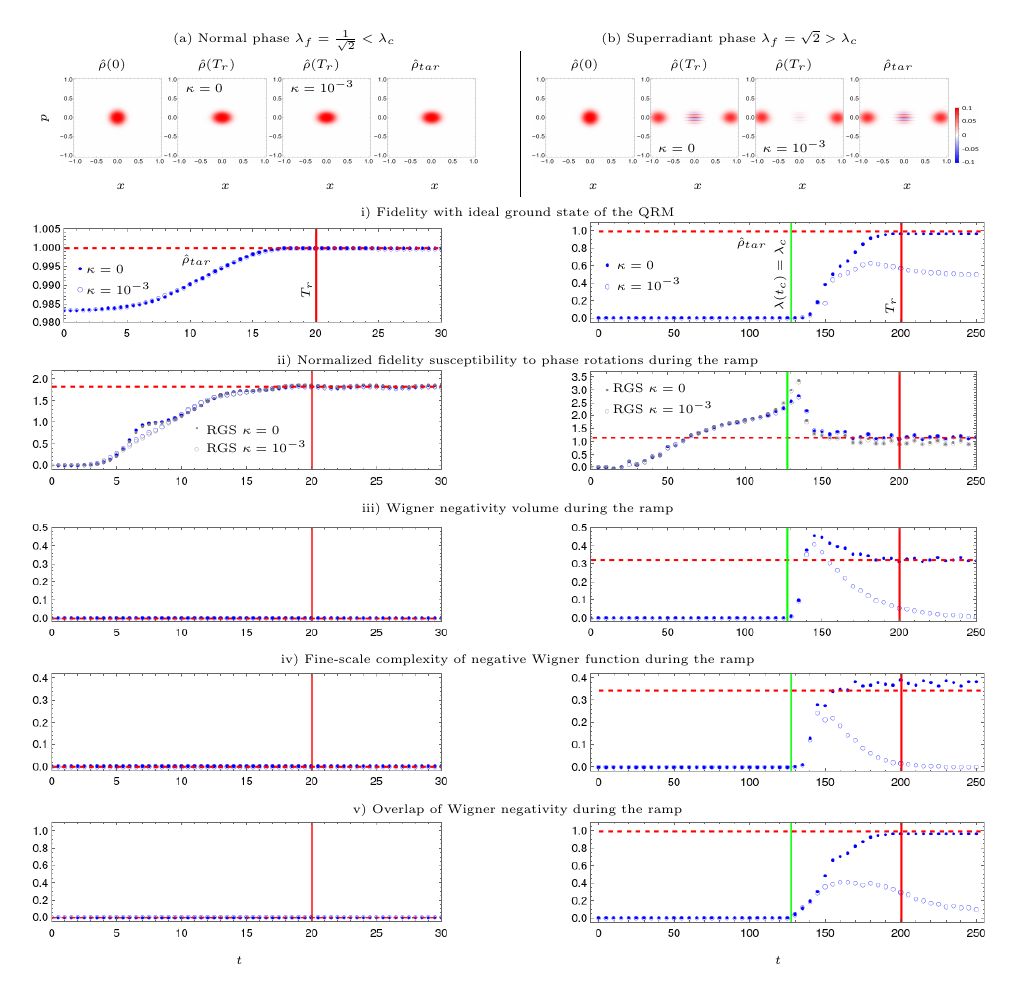}
    \caption{\label{ramp2} Ramp protocol to approach the ground state of the QRM. (a) The final coupling strength  remains within the normal phase $\lambda_f=\frac{1}{\sqrt{2}}<\lambda_c$ and the ramp period is short $T_r=20$. (b) The final coupling strength crosses into the superradiant phase $\lambda_f=\sqrt{2}>\lambda_c$ and $T_r=200$.
i) Fidelity with respect to the target state ${\mathcal F}(\hat{\rho}(t),\hat{\rho}_{tar})$. ii) Normalized fidelity susceptibility to phase rotations $\bar{F}_s[\hat{\rho}_b(t),\hat{n}]$. iii) Wigner negativity volume $\delta(t)$. iv) Fine-scale complexity of Wigner negativity $\chi^2_-(t)$. v) Overlap of Wigner negativity $N(t)$ with respect to the target ground state. All quantifiers are presented as a function of time $t$. The red horizontal dashed line corresponds to the target ground state $\hat{\rho}_{tar}=|E_0(\lambda_f)\rangle\langle E_0(\lambda_f)|$. For all figures, filled circles denote the evolution without loss $\kappa=0$, whereas empty circles correspond to evolution with loss $\kappa=10^{-3}$. The green line corresponds to the time when the ramp crosses the critical point $\lambda(t_c)=\lambda_c$ and the red line is the ramp durations $T_r=20$ and $T_r=200$ for the ramp to the normal and superradiant phase, respectively. The qubit frequency is $R=50$.
    }
\end{figure*}

 The spectral properties of the QRM Hamiltonian \eqref{HQRM} determine the adiabatic timescales, since the excitation probability scales with the energy gap between the ground and first excited states as $P_{\rm ex}\propto |E_0-E_1|^{-4}$ \cite{Polkovnikov2008}. In the normal phase $\lambda_f<\lambda_c$, the energy gap $|E_0-E_1|$ remains finite, allowing relatively short adiabatic ramps. In the superradiant phase $\lambda_f>\lambda_c$, the gap closes exponentially \cite{PhysRevLett.115.180404}, requiring much longer adiabatic ramps. In Fig. \ref{ramp2}, we simulate two representative cases: one with a final coupling strength within the normal phase $\lambda_f=\frac{1}{\sqrt{2}}<\lambda_c$, and one in which the coupling strength crosses into the superradiant phase $\lambda_f=\sqrt{2}>\lambda_c$. The first case corresponds to step 1 of the protocol proposed in this work. In this case (a), the ground state of the QRM is reached with high fidelity over a relatively short ramp period $T_r=20$. The normalized fidelity susceptibility to phase rotations increases together with that of its reference Gaussian state because the state is only squeezed. Both fidelity and its susceptibility are robust against loss. However, all quantifiers of quantum non-Gaussian effects are zero as the state remains Gaussian in the normal phase. In the second case (b), a significantly longer ramp period $T_r=200$ is required to reach the ground state of the QRM with reasonable fidelity, but the fidelity is very sensitive to loss.  The normalized fidelity susceptibility to phase rotations is also enhanced from its initial value, reaches a maximum close to the time where the coupling crosses the critical point and then decays. The normalized fidelity susceptibility of the generated state and that of the reference Gaussian state, having the same covariance matrix, go asymptotically to different values, indicating that the state is not Gaussian once the coupling strength crosses the critical point. The Wigner negativity volume and the complexity of the negative Wigner function exhibit a sudden increase once the coupling strength crosses the critical point. This effect represents a directly measurable signature of criticality in the QRM and will be analyzed elsewhere. After the critical point, the negativity volume and the complexity of the negative Wigner function approach the value of the target state. The negativity overlap also exhibits a sudden increase once the coupling strength crosses the critical point and approaches unity. However, when weak loss is considered $\kappa=10^{-3}$, both the fidelity and the quantifiers of quantum non-Gaussian effects are significantly degraded with respect to those of the ideal ground state of the QRM. After the ramp, in the presence of loss, the quantum non-Gaussian effects 
 decay to a small fraction of their lossless values, whereas those produced by resonant modulation in Fig.~\ref{example_protocol} remain robust over the whole simulated window, even though the modulation ramps are longer. This indicates that the quantum non-Gaussian features of the QRM ground state can be fragile against decoherence when the coupling strength crosses into the superradiant phase.

\section{Classical resonances under modulation}\label{resonant}

Repeating the semiclassical arguments in Appendix \ref{periodsapp}, one can derive time-periodic semiclassical Hamiltonians corresponding to the QRM with continuous modulation \eqref{HdrivenQRM}
{\footnotesize
\begin{eqnarray}
\label{Hsemiclassicalperiodic}
h_{eff}^{(\Vec{\eta}=\Vec{z})}(t)&=&\frac{p^2+x^2}{2}\pm \frac{1}{2}\sqrt{(1+\xi\cos(\frac{2\pi}{\tau} t))^2 + 2\lambda^2x^2}\,, \nonumber \\
h_{eff}^{(\Vec{\eta}=\Vec{x})}(t)&=&\frac{p^2+x^2}{2}\pm \frac{1}{2}\sqrt{1 + \left(\sqrt{2}\lambda x+\xi\cos(\frac{2\pi}{\tau} t)\right)^2}\, .
\end{eqnarray}}
Notice that in the parity-conserving  modulation $\Vec{\eta}=\Vec{z}$, the time-periodic semiclassical Hamiltonian $h_{\rm eff}^{(\hat{\eta}=\hat{z})}(t)$ remains invariant under the transformation $x\to -x$ and $p\to -p$. For the parity-broken modulation $\Vec{\eta}=\Vec{x}$, the time-periodic semiclassical Hamiltonian $h_{\rm eff}^{(\Vec{\eta}=\Vec{x})}(t)$ preserves only the symmetry under the transformation $p\to -p$.

The classical phase space picture corresponding to the effective time-periodic Hamiltonians $h_{eff}(t)$ can be visualized through the Poincar\'e maps \cite{poncairemaps}. These maps are constructed by evolving a set of initial conditions according to the canonical Hamilton equations and recording their phase space positions once every modulation period. For small modulation amplitudes, $\xi \ll 1$, the Poincar\'e maps are expected to resemble the trajectories of the QRM without modulation, following the Kolmogorov--Arnold--Moser (KAM) theorem \cite{Lichtenberg1992RegularAC}, which guarantees the persistence of most nonresonant invariant tori under sufficiently weak perturbations.

\begin{figure}[t!]
    \centering
    \includegraphics[width=\columnwidth]{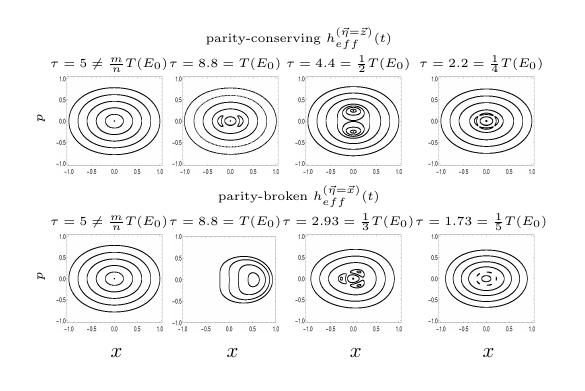}
    \caption{\label{Poincare} Poincar\'e maps corresponding to the semiclassical time-dependent Hamiltonians \eqref{Hsemiclassicalperiodic}. The top panel corresponds to the parity-conserving case $h_{eff}^{(\Vec{\eta}=\Vec{z})}(t)$ and the bottom panel to the parity-broken  case $h_{eff}^{(\Vec{\eta}=\Vec{x})}(t)$. The modulation periods are selected to match the resonant conditions $\tau=\frac{1}{n}T(E_0)$, with $n$ an integer and the characteristic period  $T(E_0)=8.8$. The coupling strength lies below the critical coupling $\lambda=\frac{1}{\sqrt{2}}<\lambda_c$. For all figures,  the modulation amplitude $\xi=\frac{1}{20}$.
    }
\end{figure}

However, invariant tori associated with rational (resonant) frequency ratios are not protected by the KAM theorem. Instead, the Poincar\'e--Birkhoff theorem \cite{birkhoff1926extension} predicts that invariant tori whose periods satisfy the resonant condition
\begin{equation}
\label{resonances}
\frac{\tau}{T(E)} = \frac{m}{n},
\end{equation}
are destroyed even for arbitrarily weak modulation amplitudes \cite{Lichtenberg1992RegularAC,KAMresonances}. Here, $T(E)$ denotes the period of classical trajectories of the unperturbed Hamiltonian (\ref{classicalT}) at energy $E$, $\tau$ is the modulation period, and $m,n$ are 
positive integers. Here, $m/n$ is not reduced to lowest terms: the resonant term is proportional
to $\cos(n\theta-2\pi m t/\tau)$, with $\theta$ the orbital angle, and $n$ equals
the number of islands in the resulting chain. Since the potential in~\eqref{Hsemiclassicalperiodic} is even, $x(\theta)$ contains only odd and $x^2(\theta)$ only even harmonics.
Hence, the parity-conserving modulation, entering~\eqref{Hsemiclassicalperiodic} only through $x^2$,
produces resonances with even $n$, whereas the parity-broken modulation,
coupling linearly to $x$, produces resonances with odd $n$. In particular, the
resonance at $\tau\approx T(E_0)$ in the parity-conserving case corresponds to
$m/n=2/2$; it arises at second order in $\xi$ and appears as a two-island chain
(Fig.~\ref{Poincare}).
As a consequence, the resonant torus, together with nearby invariant tori, is reorganized into a chain of $n$ islands, leading to a qualitative restructuring of the classical phase space. The width of the islands declines rapidly with increasing $n$. The Poincar\'e maps for resonances near the origin of the phase space are shown in Fig.~\ref{Poincare}.

\section{Floquet states near resonances}\label{Floquetapp}

\begin{figure}[t!]
    \centering
    \includegraphics[width=\columnwidth]{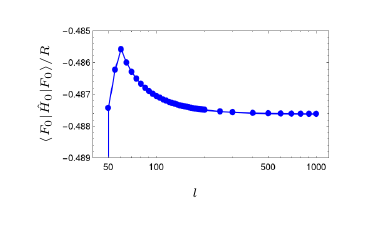}
    \caption{\label{convergencel} Mean energy $\langle F_0 |\hat{H}_0|F_0\rangle/R$ in the Floquet eigenstate $|F_0\rangle$ as a function of the number of time subintervals $l$ (in log scale).  For the analysis, we use a coupling strength below the critical point $\lambda=\frac{1}{\sqrt{2}}<\lambda_c$ and qubit frequency $R=50$. The qubit modulation preserves parity $\hat{\eta}=\hat{z}$, the amplitude is $\xi=\frac{1}{20}$ and the modulation period is near resonance $\tau=4.4=\frac{1}{2}T(E_0)$.
    }
\end{figure}

The time-periodic Hamiltonian with continuous modulation (\ref{HdrivenQRM}) can conveniently be written as $\hat{H}=\hat{H}_0+\hat{V}(t)$, where $\hat{H}_0$ is the QRM and $V(t)=\xi\frac{R}{2}\hat{\eta}\cdot\Vec{\sigma}\cos(\frac{2\pi}{\tau} t)$ is the qubit modulation, in order to perform the Trotter-Suzuki decomposition \cite{trotter1959product,Hatano2005,suzuki1985decomposition}. According to the Floquet theory \cite{floquet1883equations} for periodic time-dependent Hamiltonians like (\ref{HdrivenQRM}), the unitary dynamics over an integer number of periods is formally solved by $|\Psi(t=n\tau)\rangle =\hat{F}^n(0,\tau)|\Psi(0)\rangle$, where $\hat{F}(0,\tau)$ is the Floquet operator \cite{GRIFONI1998229,LAPTYEVA201685}. Just as the unitary dynamics of a system governed by a time-independent Hamiltonian can be described in terms of its energy eigenstates, the unitary dynamics of a time-periodic system can be described in terms of the eigenstates $|F_k\rangle$ of the Floquet operator $\hat{F}(0,\tau)$ \cite{verdeny2016quasi}. The states $|F_k\rangle$, known as Floquet states, are quasistationary over the modulation period and can exhibit topological properties and statistical distributions that differ from those of eigenstates of static Hamiltonians \cite{Mori_Floquet,d2014long}. The Floquet operator corresponding to the time-dependent Hamiltonian (\ref{HdrivenQRM}) can be approximately obtained using the Trotter-Suzuki decomposition \cite{Hatano2005}
{\small
\begin{eqnarray}
\label{TrotterSuzuki}
\hat{F}(0,\tau)&\approx& \exp\left(-i\hat{H}_0\frac{\Delta t}{2}\right) \exp\left( -i\int^{l\Delta t}_{(l-1)\Delta t} \hat{V}(t)dt \right)\exp\left(-i\hat{H}_0\frac{\Delta t}{2}\right) \nonumber \\
      & & ... \nonumber \\
      & & \exp\left(-i\hat{H}_0\frac{\Delta t}{2}\right) \exp\left( -i\int^{2\Delta t}_{\Delta t} \hat{V}(t)dt \right)\exp\left(-i\hat{H}_0\frac{\Delta t}{2}\right) \nonumber \\
      & &\exp\left(-i\hat{H}_0\frac{\Delta t}{2}\right) \exp\left( -i\int^{\Delta t}_{0} \hat{V}(t)dt \right)\exp\left(-i\hat{H}_0\frac{\Delta t}{2}\right)
\end{eqnarray}
}
where $l$ is an integer and $\Delta t=\frac{\tau}{l}$ is a portion of the modulation period. In Fig. \ref{convergencel} we show the convergence of the Floquet eigenstate $|F_0\rangle$ as $l$ increases.


In the weak-modulation regime $\xi \ll 1$, the Floquet states can be conveniently sorted according to the expectation value $\langle F_k \lvert \hat{H}_0 \rvert F_k \rangle$, such that the subindex $k$ indicates the stationary state $|E_k\rangle$ to which it is adiabatically connected~\cite{KAMresonances,segura2025quantum}.

\begin{figure}[t!]
    \centering
    \includegraphics[width=\columnwidth]{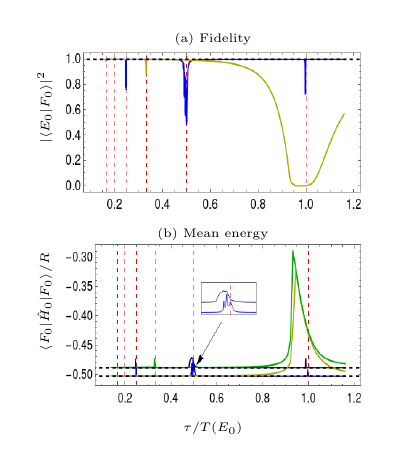}
    \caption{\label{fidelity_exp}
     (a)  Fidelity $|\langle E_0|F_0 \rangle|^2$ between the lowest Floquet state and the ground state of the  QRM. The deviations from unity illustrate the contribution from  higher states of the QRM in the resonant condition. (b) Mean energy given by the expectation value $\langle F_0 |\hat{H}_0|F_0\rangle/R$ in the lowest Floquet state $|F_0\rangle$.  The blue line corresponds to the case of parity-conserving  modulation $\Vec{\eta}=\Vec{z}$ and the green line to the case of parity-broken  modulation $\Vec{\eta}=\Vec{x}$. The black dashed lines indicate the results for vanishing modulation amplitude $\xi=0$ where $|E_0\rangle=|F_0\rangle$. The red vertical lines indicate the resonant conditions $\tau=\frac{1}{n}T(E_0)$ with $n=1,2,3,4,5,6$. The darker colors correspond to the second lowest Floquet state $|F_1\rangle$. Crossings (or avoided crossings) between Floquet states indicate that adiabatic continuation with respect to a single eigenstate of the QRM Hamiltonian is not guaranteed.
     For all the figures, the modulation amplitude was selected at $\xi=\frac{1}{20}$, the coupling strength within the normal phase $\lambda=\frac{1}{\sqrt{2}}<\lambda_c$ and the qubit frequency $R=50$.
    }
\end{figure}

 There is growing evidence that the Floquet states $|F_k\rangle$ of time-periodic Hamiltonians reflect the classical resonances (\ref{resonances}) in their phase space representations. Pronounced restructuring of Floquet eigenstates has been reported near resonances in kicked oscillators~\cite{billam2009quantum,PhysRevE.109.014209}, the Harper model~\cite{wisniacki}, stroboscopic maps~\cite{wisniacki2015quantum}, and a variety of driven spin systems~\cite{segura2025quantum,CauxPRL2018,PhysRevB.105.174205}. As a first step in the search for resonant effects, we examine the fidelity between the lowest Floquet state and the ground state of the QRM $|\langle F_0 | E_0 \rangle|^2$, together with the expectation value of the time-independent Hamiltonian evaluated in the Floquet state $\langle F_0 | \hat{H}_0 | F_0 \rangle$, both as functions of the normalized modulation period $\tau/T(E_0)$. To isolate resonant effects, we focus on a weak modulation amplitude $\xi = 1/20$, and vary the modulation period up to the characteristic period of the ground state of the QRM $T(E_0) \approx 8.8$, as defined in (\ref{quantumT}). The coupling strength is chosen within the normal phase $\lambda = 1/\sqrt{2}$ and the qubit frequency is finite $R = 50$. The results shown in Fig.~\ref{fidelity_exp} indicate that the Floquet and ground states remain very close over a wide range of modulation periods, with the fidelity remaining close to unity and deviating significantly only in the vicinity of the resonant conditions $\tau \approx \frac{m}{n}T(E_0)$. Importantly, in the parity-conserving /broken modulation $\Vec{\eta} = \Vec{z},\Vec{x}$, only resonances with even/odd $n$ are observed, respectively. Consistently, the expectation value $\langle F_0 | \hat{H}_0 | F_0 \rangle$ remains also close to the ground-state energy $E_0$ and exhibits significant variations only near the resonances. Together with the fidelity, this behavior provides clear evidence of the adiabatic continuation between the QRM ground state $|E_0\rangle$ and the lowest Floquet state $|F_0\rangle$. Very close to crossings (or avoided crossings) between Floquet states, however, the adiabatic connection may be compromised.

\begin{figure}[t!]
    \centering
    \includegraphics[width=\columnwidth]{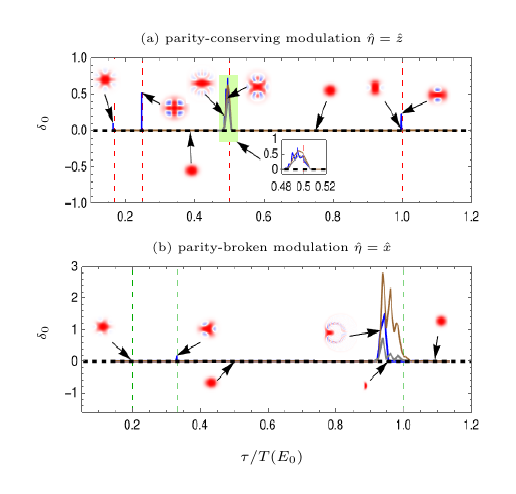}
    \caption{\label{nonGaussianstructures}
     Wigner negativity volume $\delta_0$ of the lowest Floquet state $|F_0\rangle$ as a function of the normalized modulation period $\tau/T(E_0)$. The blue line corresponds to the ideal Floquet state $|F_0\rangle$, the brown line to the state $\hat{\rho}(t=T_r)$ prepared with the ramp protocol after $T_r=80\tau$ without loss and the gray line to the state $\hat{\rho}(t=T_r)$ prepared with the ramp protocol after $T_r=80\tau$ with bosonic loss $\kappa=10^{-3}$.
     The top panel (a) shows the parity-conserving  modulation $\Vec{\eta}=\Vec{z}$ and the bottom panel (b) shows the parity-broken  modulation $\Vec{\eta}=\Vec{x}$. For both panels, the modulation amplitude is fixed $\xi=1/20$, the coupling strength $\lambda=1/\sqrt{2}<\lambda_c$ is within the normal phase and the qubit frequency is finite $R=50$. The dashed black line corresponds to the Wigner negativity volume in the case without modulation. The insets correspond to the Wigner function of the ideal Floquet states at the specific modulation periods. 
     The vertical dashed red lines correspond to the even resonances $\tau=\frac{m}{n}T(E_0)$ with $m/n=1/6,1/4,1/2,2/2$, and the green lines to the odd resonances with $m/n=1/5,1/3,1/1$.
    }
\end{figure}

The emergence of quantum non-Gaussian features in the Floquet state $|F_0\rangle$ near the Poincaré--Birkhoff resonances is illustrated in Fig.~\ref{nonGaussianstructures} through the Wigner negativity volume $\delta_0$ as a function of the normalized modulation period $\tau/T(E_0)$. Depending on the modulation direction $\vec{\eta}$ and the modulation period $\tau$, the Wigner function of the Floquet state $|F_0\rangle$ develops a rich variety of quantum non-Gaussian structures. For comparison, we also show the Wigner negativity of the states generated through the adiabatic-ramp protocol, both in the absence and in the presence of loss.

\begin{figure*}[!htbp]
    \centering
    \includegraphics[width=\textwidth]{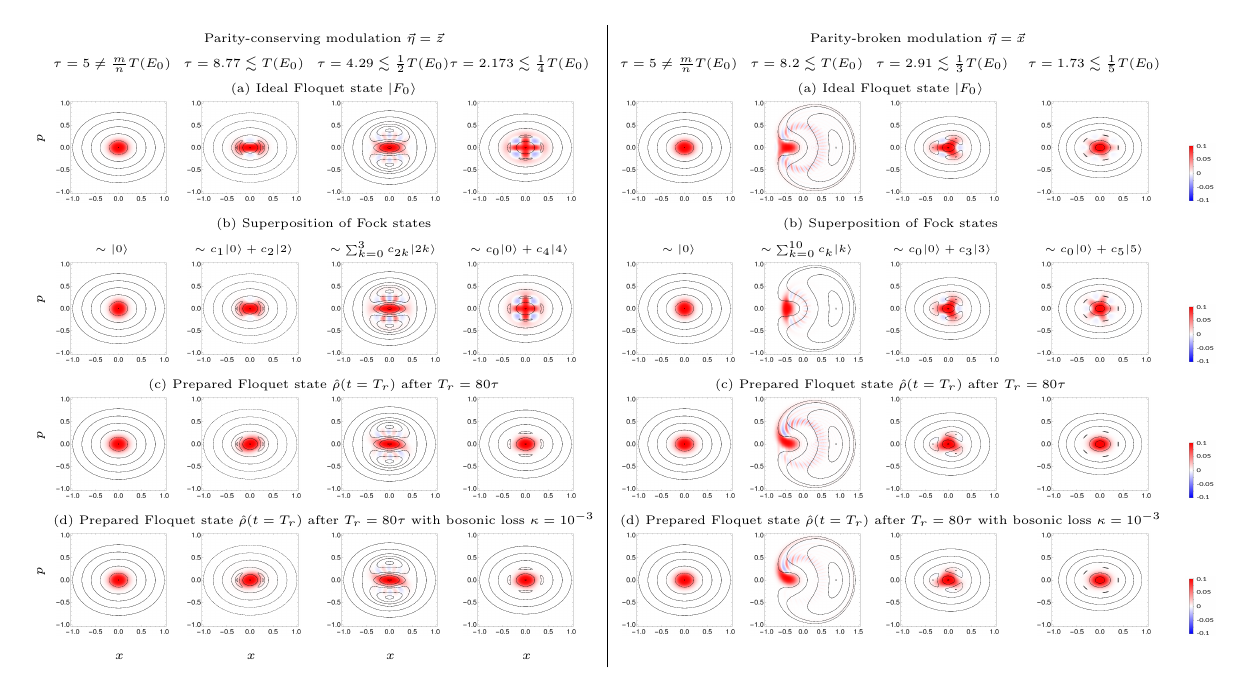}
    \caption{\label{resonantcat} Wigner function of the lowest Floquet state $|F_0\rangle$ at specific qubit modulation periods close to the resonances $\tau\lesssim\frac{1}{n}T(E_0)$. The comparison includes (a) ideal Floquet states $\hat{\rho}_{tar}=|F_0\rangle\langle F_0|$, (b) superposition of Fock states that the Floquet state resembles (c) Floquet states $\hat{\rho}(t=T_r)$ prepared with the ramp protocol after $T_r=80\tau$ and (d) Floquet states $\hat{\rho}(t=T_r)$ prepared with the ramp protocol after $T_r=80\tau$ with bosonic loss $\kappa=10^{-3}$.  In the background of each Wigner function, we include the corresponding Poincar\'e maps from the time-periodic semiclassical Hamiltonians \eqref{Hsemiclassicalperiodic}.
    The coupling strength lies within the normal phase $\lambda=\frac{1}{\sqrt{2}}<\lambda_c$ for which the characteristic period is $T(E_0)=8.8$. For all figures, the qubit frequency is finite $R=50$ and the modulation amplitude is fixed $\xi=1/20$.
    }
\end{figure*}

In Fig.~\ref{resonantcat}, the Wigner functions of the lowest Floquet state $|F_0\rangle$ near different Poincaré--Birkhoff resonances are compared with the corresponding semiclassical Poincar\'e maps. The comparison makes evident that the quantum non-Gaussian features of the Floquet state $|F_0\rangle$ are a consequence of its coherent delocalization over a region of phase space containing classically disconnected resonance islands. Near the resonant conditions, the Floquet state $|F_0\rangle$ acquires structures resembling a superposition of a small number of Fock states. 
 In particular, $|F_0\rangle\approx c_0|E_0\rangle+c_n|E_n\rangle$ near the resonance $m/n$, whose Wigner function has the $n$-fold symmetry of the classical island chain.
For parity-conserving  modulation, $\vec{\eta}=\vec{z}$, the Floquet state contains the vacuum and even Fock states with even components, whereas for parity-broken  modulation, $\vec{\eta}=\vec{x}$, states with odd parity contribute. However, within the finite ramp duration considered here $T_r=80\tau$, the target Floquet state $|F_0\rangle$ can be efficiently approached only for resonances $\frac{1}{n}T(E_0)$ with $n=1$, $n=2$ and only marginally $n=3$. Other resonances could be accessed by increasing the qubit frequency $R$ towards the semiclassical limit. Another alternative is to decrease the adiabatic preparation times with shortcut techniques (see \cite{guery,Jarzynski,Berry_2009,Chen2021} for instance).

\end{document}